\documentclass[preprint,showpacs,preprintnumbers,amsmath,amssymb]{revtex4}

\usepackage{graphicx}
\usepackage{dcolumn}
\usepackage{bm}

\begin{document}


\title{Transmission area and two-photon correlated imaging}

\author{Yanfeng Bai}
\author{Shensheng Han}
\author{Honglin Liu}%
\affiliation{Key Laboratory for Quantum Optics and Center for Cold
Atom Physics, Shanghai Institute of Optics and Fine Mechanics,
Chinese Academy of Sciences, Shanghai 201800, China
}%

\date{\today}

\begin{abstract}
The relationship between transmission area of an object imaged and
the visibility of its image is investigated in a lensless system.
We show that the changes of the visibility are quite different
when the transmission area is varied by different manners. An
increase of the transmission by adding the slit number leads to a
decrease of the visibility. While, the change is adverse when the
slit width is widened for a given distance between two slits.
\end{abstract}

\maketitle

The topic of two-photon correlated imaging has attracted much
attention in recent
years\cite{Ribeiro94,Belinsky94,Pittman95,Strekalov95,Abouraddy02,Gatti99}.
The theory and experiment of ghost imaging with two-photon quantum
entanglement were firstly demonstrated in
mid-1990s\cite{Belinsky94,Pittman95}. A fourth-order correlation
function in the $\{{\bf r}\}$ space is developed and applied to
double-slit experiments with spontaneous down-conversion light by
Barbosa\cite{Barbosa96}. Using the coupling between polarization
entanglement and the entanglement for the transverse degrees of
freedom, Caetano {\em et al.} demonstrated both theoretically and
experimentally the manipulation of quantum entangled
images\cite{caetano03}. The role of entanglement in two-photon
imaging was discussed, Abouraddy {\em et al.} showed that
entanglement is a prerequisite for achieving distributed quantum
imaging\cite{Abouraddy01}. While their work leads to some debate.
Bennink {\em et al.} showed that coincidence imaging does not
require entanglement, and provided an experimental demonstration
using a classical source\cite{Bennink04}. Using classical
statistical optics, Cheng and Han studied a particular aspect of
coincidence imaging with incoherent sources\cite{Cheng04}. The
first experimental demonstration of two-photon correlated imaging
with true thermal light from a hollow cathode lamp was also
reported\cite{Zhang05}.

Many objects imaged have been chosen since the theory of
coincidence imaging was proposed. A double-slit was usually
selected as an object in two-photon correlated
imaging\cite{Gatti99,Barbosa96,Bennink04,Cheng04,Zhang05,Gatti04,Choi99,Magatti04,Jing05}.
Cai and Zhu designed an optical system for implementing the
second-order fractional Fourier transform for a single
slit\cite{Cai05}. Eisebitt {\em et al.} demonstrated a versatile
approach to perform lensless imaging of a sample consisting of a
``H", an ``I", and an open square at x-ray
wavelength\cite{Eisebitt04}.

\begin{figure}
\includegraphics[width=6cm]{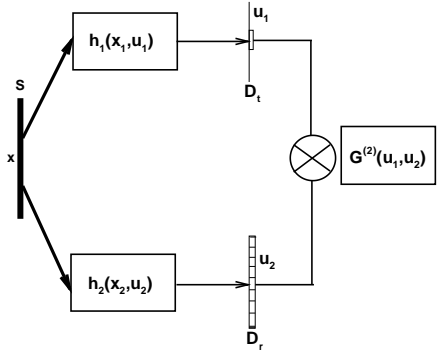}
\caption{\label{fig1} Two-photon correlated imaging with
incoherent light.}
\end{figure}

In this paper, we investigate the effects from transmission area
of an object on the coincidence imaging with incoherent light
source. Here transmission area is varied by two manners,
increasing the slit number and slit width. We show that the
effects are quite different under two conditions. Here the imaging
system appropriate for correlated imaging is shown in
Fig.~\ref{fig1}. The light beam from the source $S$ is incoherent,
and divided into two beams by a beam splitter, they travel through
a test and a reference arms, which are described by their impulse
response functions $h_{1}(x_{1},u_{1})$ and $h_{2}(x_{2},u_{2})$,
respectively. The test arm usually includes an object to be
imaged. Detector $D_{t}$ is a pointlike detector, $D_{r}$ is an
array of pixel detector, which are used to record the intensity
distribution of the photons at $u_{1}$ and $u_{2}$, respectively.

The optical field in the source can be represented by $E(x)$.
After propagating through two different optical systems, the field
has

\begin{equation}{\label{field}}
E_{i}(u_{i})=\int E(x)h_{i}(x,u_{i})dx,\hspace{2mm}i=1,2.
\end{equation}

The fourth-order correlation function at $u_{1}$ and $u_{2}$ may
be recorded with the coincidence rate in the test and reference
detectors

\begin{equation}\label{coin}
G^{(2)}(u_{1},u_{2})=\langle
E(u_{1})E(u_{2})E^{*}(u_{2})E^{*}(u_{1})\rangle,
\end{equation}
where $E(u_{i})$ $(i=1,2)$ is the optical field in the test
(reference) detector. Substituting Eq.~(\ref{field}) into
Eq.~(\ref{coin}), we have

\begin{eqnarray}\label{xcoin}
G^{(2)}(u_{1},u_{2})&=&\int\int\int\int\langle
E(x_{1})E(x_{2})E^{*}(x'_{2})E^{*}(x'_{1})\rangle\nonumber\\
&&\times
h_{1}(x_{1},u_{1})h_{2}(x_{2},u_{2})h^{*}_{2}(x'_{2},u_{2})\nonumber\\
&&\times h^{*}_{1}(x'_{1},u_{1})dx_{1}dx_{2}dx'_{2}dx'_{1},
\end{eqnarray}
where $\langle E(x_{1})E(x_{2})E^{*}(x'_{2})E^{*}(x'_{1})\rangle$
is the four-order correlation function at the light source, We
represent it by $G^{(2)}(x_{1},x_{2},x'_{2},x'_{1})$ in the
following in order to outline the parallelism with the formalism
in Eq.~(\ref{coin}).

In many cases, the fluctuation of a classical light field can be
characterized by a Gaussian field statistics with zero
mean\cite{Goodman}, one obtains

\begin{eqnarray}\label{sec}
G^{(2)}(x_{1},x_{2},x'_{2},x'_{1})&=&G^{(1)}(x_{1},x'_{1})G^{(1)}(x_{2},x'_{2})\nonumber\\
&&+G^{(1)}(x_{1},x'_{2})G^{(1)}(x_{2},x'_{1}),
\end{eqnarray}
where $G^{(1)}(x_{i},x_{j})$ is the second-order correlation
function of the fluctuating source field, and arbitrary order
correlation function is thus expressed via the second-order
correlation function

\begin{equation}\label{arb}
G^{(1)}(x_{i},x_{j})=\langle E(x_{i})E^{*}(x_{j})\rangle,
\end{equation}
along with the relation
$G^{(1)}(x_{i},x_{j})=[G^{(1)}(x_{j},x_{i})]^{*}$. By using
Eqs.~(\ref{sec})and (\ref{arb}), we can simplify Eq.~(\ref{xcoin})
as

\begin{widetext}
\begin{eqnarray}\label{zui}
G^{(2)}(u_{1},u_{2})&=&\int\int\langle
E(x_{1})E^{*}(x'_{1})\rangle
h_{1}(x_{1},u_{1})h^{*}_{1}(x'_{1},u_{1})dx_{1}dx'_{1}\nonumber\\
&&\times\int\int\langle E(x_{2})E^{*}(x'_{2})\rangle
h_{2}(x_{2},u_{2})h^{*}_{2}(x'_{2},u_{2})dx_{2}dx'_{2}\nonumber\\
&&+\int\int\langle E(x_{1})E^{*}(x'_{2})\rangle
h_{1}(x_{1},u_{1})h^{*}_{2}(x'_{2},u_{2})dx_{1}dx'_{2}\nonumber\\
&&\times\int\int\langle E(x_{2})E^{*}(x'_{1})\rangle
h_{2}(x_{2},u_{2})h^{*}_{1}(x'_{1},u_{1})dx_{2}dx'_{1}\nonumber\\
&=&\langle I(u_{1})\rangle\langle I(u_{2})\rangle+\Big|\int\int
\langle E(x_{1})E^{*}(x'_{2})\rangle
h_{1}(x_{1},u_{1})h^{*}_{1}(x'_{2},u_{2})dx_{1}dx'_{2}\Big|^{2}\nonumber\\
&=&\langle I(u_{1})\rangle\langle I(u_{2})\rangle+\Delta
G^{(2)}(u_{1},u_{2}).
\end{eqnarray}
\end{widetext}

The information of the object imaged is extracted by measuring the
spatial correlation function of the intensities $\langle
I(u_{1})I(u_{2})\rangle$. By subtracting the background term
$\langle I(u_{1})\rangle\langle I(u_{2})\rangle$, we can obtain
the correlation function of intensity fluctuations, all
information about the object is contained in it

\begin{equation}\label{flu}
\Delta G^{(2)}(u_{1},u_{2})=\langle
I(u_{1})I(u_{2})\rangle-\langle I(u_{1})\rangle\langle
I(u_{2})\rangle,
\end{equation}

Theoretically the visibility which is our concern in this paper is
defined as

\begin{eqnarray}
V&=&\frac{\Delta G^{(2)}(u_{1},u_{2})_{max}}{\langle
I(u_{1})I(u_{2})\rangle_{max}}\nonumber\\
&=&\frac{\Delta G^{(2)}(u_{1},u_{2})_{max}}{\Big[\langle
I(u_{1})\rangle\langle I(u_{2})\rangle+\Delta
G^{(2)}(u_{1},u_{2})\Big]_{max}}.
\end{eqnarray}

It has been known that, from the discussion in
Ref.~\cite{Bache06}, $\Delta G^{(2)}\leq\langle
I(u_{1})\rangle\langle I(u_{2})\rangle$ in the thermal case, so
the visibility is never above $0.5$. In the following, we will
investigate the effects from the transmission area of an object on
the visibility.

For incoherent light, we assume that its intensity distribution is
of the Gaussian type. Then the two-order correlation function for
completely incoherent light source can be written as

\begin{equation}\label{source}
\langle
E_{s}(x_{1})E_{s}^{*}(x_{2})\rangle=G_{0}\exp\Big(-\frac{x_{1}^{2}+x_{2}^{2}}{4a^{2}}\Big)\delta(x_{1}-x_{2}),
\end{equation}
where $G_{0}$ is a normalized constant, $a$ is the transverse size
of the source.

In the test arm, an object (transmission function $t(x')$) is
located at a distance $z_{1}$ from the source $S$ and a distance
$z_{2}$ from detector $D_{t}$. Thus the impulse response function
can be expressed as

\begin{eqnarray}\label{w1}
h_{1}(x_{1},u_{1})&=&\int
\frac{e^{-ikz_{1}}}{i\lambda z_{1}}\exp\Big[\frac{-i\pi}{\lambda z_{1}}(x'-x_{1})^{2}\Big]t(x')\nonumber\\
&&\times\frac{e^{-ikz_{2}}}{i\lambda
z_{2}}\exp\Big[\frac{-i\pi}{\lambda z_{2}}(u_{1}-x')^{2}\Big]dx'.
\end{eqnarray}

The reference arm contains nothing but free-space propagation from
$S$ to $D_{r}$. Thus the corresponding impulse response function
under the paraxial approximation is
\begin{equation}\label{w2}
h_{2}(x_{2},u_{2})=\frac{e^{-ikz}}{i\lambda
z}\exp\Big[\frac{-i\pi}{\lambda z}(u_{2}-x_{2})^{2}\Big],
\end{equation}
we know that, from the conclusion in Ref.~\cite{Cheng04}, such a
coincidence imaging system realizes the function of Fourier
transform imaging under the condition of a large, uniform, fully
incoherent light source. Here, we take a double slit with slit
width $\omega=0.075$mm and the distance between two slits
$d=0.15$mm as the object imaged. The transverse size of the source
$a=1$mm, other parameters are chosen as $\lambda=532$nm,
$z=175$mm, $z_{1}=75$mm, and $z_{2}=z-z_{1}$.

\begin{figure}
\includegraphics[width=7cm]{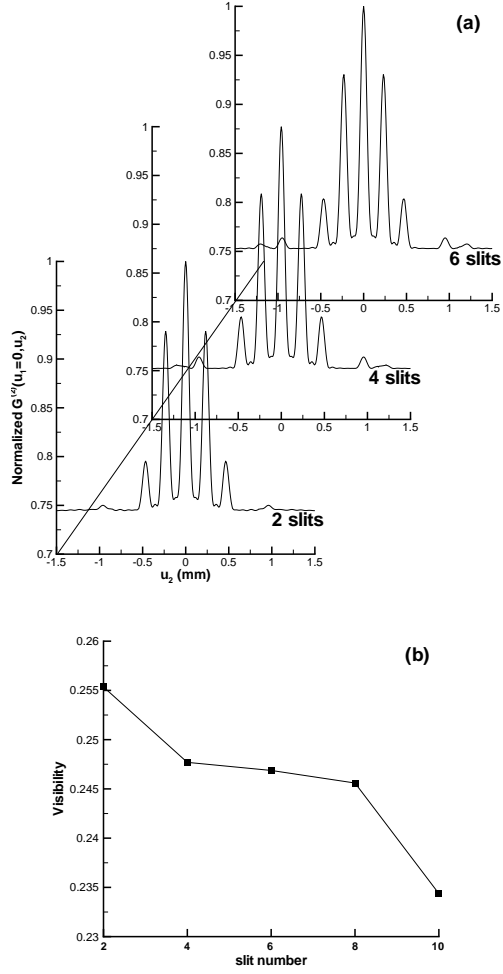}
\caption{\label{fig2} (a) The normalized conditional coincidence
rate $G^{(2)}(0,u_2)$ versus $u_{2}$ with incoherent light source
for different slit number. (b) Dependence of the corresponding
visibility on the slit number $n$.}
\end{figure}

Substituting Eqs.~(\ref{source})-(\ref{w2}) into Eq.~(\ref{zui}),
we can get the normalized conditional coincidence rate, the
numerical simulation results will be analyzed. Here, we provide
two methods to change the transmission area of the object imaged.
Firstly, we increase the slit number. The images are given in
Fig.~\ref{fig2}(a). From our simulations it clearly emerges that,
under the given parameters, the quality of the Fourier-transform
imaging gets better, whereas the visibility is decreased with an
increase of slit number, i.e., the transmission area. The
corresponding experiment results have been implemented in a
$f$-$2f$ system\cite{Liu06}.

\begin{figure}
\includegraphics[width=7cm]{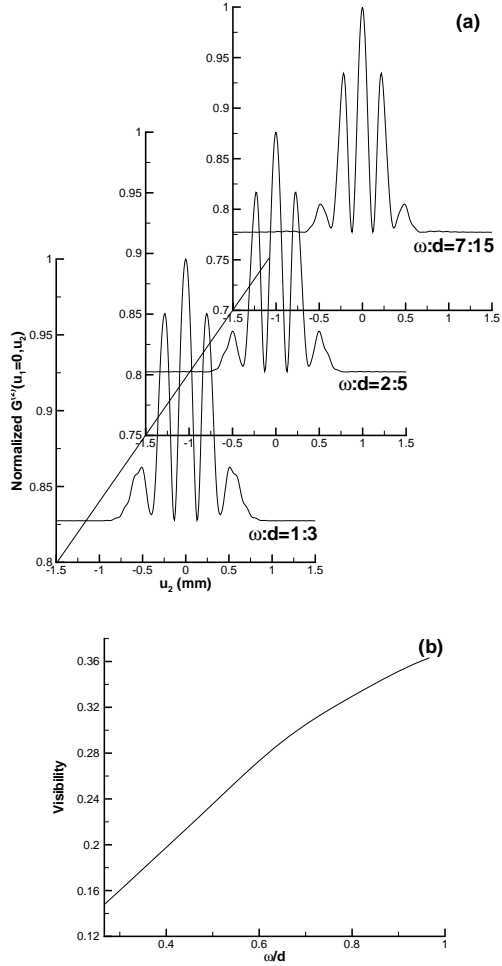}
\caption{\label{fig3} (a) The normalized $G^{(2)}(0,u_2)$ as the
function of $u_{2}$ for different ratios of the slit width to slit
distance. (b) The corresponding visibility versus $\omega/d$.}
\end{figure}

To make our results more general, in Fig.~\ref{fig2}(b), we give
the visibility under different slit number $n$. Here, we only
depict the dots for finite slits because the ghost imaging will be
distorted when $n$ is much large. It is clear that the visibility
will decrease with the increase of the slit number.

Secondly, we vary the transmission area by increasing the ratio of
the slit width to slit distance, $\omega/d$, for a given slit
distance $d=0.15$mm, the results are depicted in
Fig.~\ref{fig3}(a). An increase of the slit width leads to an
increase of the image visibility, while the image quality gets
worse during this process. It should be noticed, by comparing with
the curves in Fig.~\ref{fig2}(a), the change of the visibility by
widening the slits is much bigger than that by increasing the slit
number.

In Fig.~\ref{fig3}(b), we show the dependence of the visibility on
the normalized slit width $\omega/d$. The images are distorted
when $\omega$ is much smaller or close to $d$, so the dots should
be subtracted. Here we obtain a curve of the visibility completely
different from that in Fig.~\ref{fig2}(b). The increase of the
slit width, i.e., the transmission area, makes the visibility
enhance.

\begin{figure*}
\includegraphics[width=14cm]{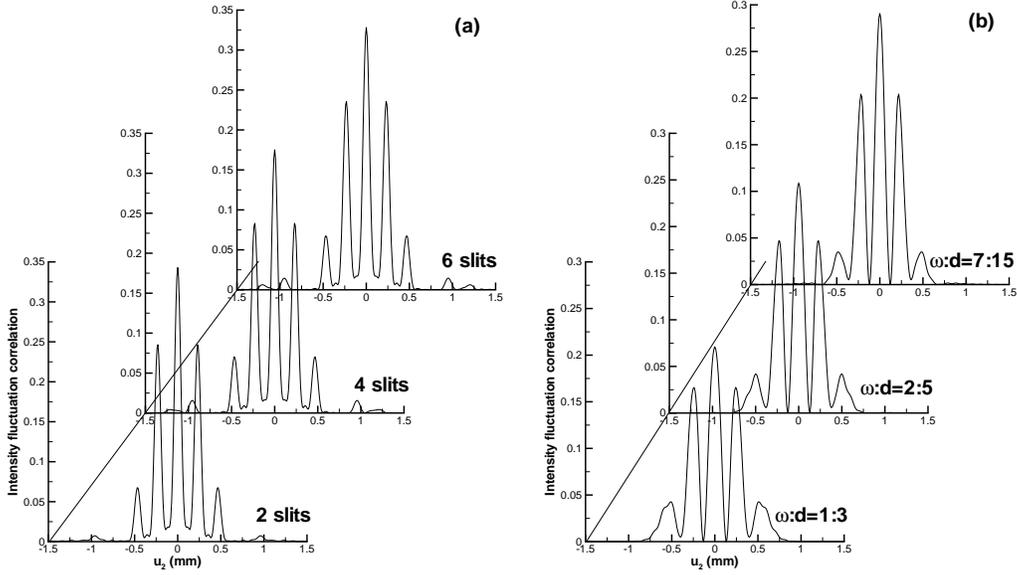}
\caption{\label{fig4} Normalized $\Delta G^{(2)}(0,u_2)$ vs the
reference detector position for an object with different
transmission. (a) Intensity fluctuation correlations for different
slit number. (b) The curves for different slit width.}
\end{figure*}

From the results in Figs.~\ref{fig2}(b) and \ref{fig3}(b), the
variation ranges of the visibility we obtain coincide with the
prediction value in Ref.~\cite{Bache06}. While the visibility is
so poor that the images can not be observed during practical
experiment implementation, so we usually retrieve the desired
information, i.e., the intensity fluctuation correlation by
subtracting the background, as shown in Fig.~\ref{fig4}. Comparing
the curves which are normalized by $\langle I(u_{1})\rangle\langle
I(u_{2})\rangle$, we can draw the same conclusion that both
decreasing the slit number and increasing the slit width make the
visibility enhance.

With the technology development of correlated imaging, one is not
satisfied with the Fourier-transform images only for simple
objects, some complex objects are being considered. Here there are
still many questions unresolved, such as what limits the
visibility for complex object. Based on the above discussions, we
can enhance the visibility of correlated imaging by choosing
proper slit number and slit width when other parameters are given.
Which can be helpful in the realistic experiments of two-photon
correlated imaging.

In conclusion, we give the theoretical analysis of the
relationship between the transmission area and the visibility in a
correlated imaging with incoherent light. Though the transmission
area can be varied by changing the slit number or width, they have
different effects on the visibility. The visibility gets worse
with an increase of the slit number, whereas an increase of the
slit width leads to an increase of the visibility.

The work was supported by the National Natural Science Foundation
of China under Grant No. 60477007, and the Shanghai Optical-Tech
Special Project (grant 034119815).

\end{document}